\documentclass[twocolumn, trackchanges]{aastex701}
\usepackage{booktabs}
\usepackage{array}
\usepackage{makecell}
\renewcommand{\arraystretch}{1.2} 
\usepackage{ragged2e}
\usepackage[hang,flushmargin]{footmisc}
\usepackage{hyperref}
\usepackage{derivative}
\usepackage{amsmath}

\def \va {v_{\text{A}}}

\begin{document}
\title{Hybrid Simulations of Supersonic Shear Flows: I) Particle Acceleration}

\author[0009-0003-0137-3116]{Naixin Liang}
\affiliation{Department of Physics, University of California, Santa Barbara, CA 93106, USA}
\email{naixin@ucsb.edu}

\author[0000-0003-0939-8775]{Damiano Caprioli}
\affiliation{Department of Astronomy \& Astrophysics, University of Chicago, Chicago, IL 60637, USA}
\affiliation{Enrico Fermi Institute, The University of Chicago, Chicago, IL 60637, USA}
\email{caprioli@uchicago.edu}

\date{\today}
\begin{abstract}\label{sec:abstract}
Supersonic flows are ubiquitous in warm and cool media; their dissipation leads to heating, generation of nonthermal particles, and amplification of background magnetic fields.
We present 2D hybrid (kinetic ions -- fluid electrons) simulations of decaying shear flows across the subsonic-to-supersonic transition, finding that the canonical Kelvin-Helmholtz instability in subsonic cases gives way to the formation of shocklets in supersonic shears, where dissipation is faster and nonthermal particles are produced.
We discuss the dependence on the flow Mach number of particle acceleration, the viscosity induced by kinetic effects, and the production of magnetic turbulence. 
We outline the potential impact of these findings for turbulence in the warm interstellar medium, for molecular clouds, and for accretion disks, leaving to a companion paper the discussion of the effects on the shear of self-generated and pre-existing energetic particles. 
\end{abstract}

\section{Introduction} \label{sec:intro}
Shearing flows are common in astrophysical environments, from relativistic jets \citep[e.g.,][]{ostrowski98, starwarz+02, rieger+04, bromberg+11, webb+18, kimura+18, rieger+19, sironi+21, tavecchio21, merten+21}, to gas motions in galaxy clusters \citep[e.g.,][]{markevitch+07, roediger+12, zuhone+16, simionescu+19}, to star-forming regions \citep[e.g.,][]{berne+12, han+19}, to stellar winds \citep[e.g.,][]{lamberts+11}, to accretion disks boundary layers \citep[e.g.,][]{belyaev+12}, to the interface between the solar wind and planetary magnetospheres \citep[e.g.,][]{johnson+14}.

Shear flows are typically prone to Kelvin--Helmholtz-like instabilities (KHI) and eventually develop into turbulence.
Conversely, turbulence intrinsically contains non-organized shear flows across a broad range of scales \citep[e.g.,][]{hunt+10,elsinga+20}. 
Turbulence pervades the interstellar medium (ISM), affecting both the structure and dynamics of interstellar gas across almost all ranges of temperature and density \citep{elmegreen+04,paredes+07}. 
In particular, ISM turbulence is seldom incompressible (subsonic) and often supersonic in cold or warm phases where temperatures are low ($T\sim 100-10^4$K) because electrons can efficiently radiate thermal energy away via semi-prohibited atomic and molecular line emission \citep[e.g.,][]{krumholz+16}. 
Supersonic turbulence in molecular clouds controls the star formation rate triggered by gas compression \citep[e.g.,][]{mckee+07} and modulates fragmentation in self-gravitating gas, affecting the initial stellar mass distribution \citep{jappsen+05,hopkins13}.
It may also have a role in controlling star formation and structure formation in the early Universe \citep[see][for a review]{federrath+13}.
As the turbulent cascade proceeds to smaller scales, the transition from supersonic to subsonic may happen at some spatial scale \citep{federrath+21}.

Very generally, shear dissipation implies a transfer of the free kinetic energy into heat, magnetic fields, and possibly nonthermal particles (henceforth \emph{cosmic rays}, CRs).
In the hydrodynamic KHI, vortices form at the interface between fluids of different velocities;
the KHI is stabilized $M > 2\sqrt{2}$ for fluids of equal density \citep{landau+44}, where $M$ is the sonic Mach number, i.e., the ratio of the bulk to sound speed.
Extension to non-constant densities is discussed by \citet{mandelker+16} in the context of cold streams in the hot medium surrounding massive galaxies. 
At high Mach numbers, in slab (or double-shear layers) geometries, a distinct regime of the instability grows in the form of supersonic \textit{body modes}, which reverberate within the slab, growing in amplitude and eventually entering a nonlinear phase characterized by shock formation and complex internal structures \citep{berlok+19b}. They may disrupt the magnetized cold streams feeding massive high redshift haloes \citep{berlok+19}.
Turbulent shears have been studied both in hydro- and magneto-hydrodynamical approaches and with kinetic simulations \citep[see, e.g.,][]{howes+08, matthaeus+11, wan+15, cerri+19, schekochihin+22, gootkin+25, achikanath-chirakkara+24, tsung+25,achikanath-chirakkara+25}, but only the latter can capture the development of the nonthermal particle distributions which may be crucial for the overall dynamics, especially for supersonic flows.


In a large-scale shear flow, particles bounce back and forth probing the velocity profile: such a shear acceleration, usually studied in the context of astrophysical jets, can be viewed as a quasi-coherent form of stochastic Fermi acceleration \citep[e.g.,][]{jokipii87, ohira+13, ostrowski98, rieger+19, webb+18}. 
Nonlinear KH vortices also produce kinetic-scale reconnection regions, which may efficiently extract particles from the thermal pool and inject them into shear acceleration \citep{sironi+21}.
Very recently \citet{liu+25} have used simulations with a magneto-hydrodynamical (MHD) thermal background and particle-in-cell (PIC) nonthermal population \citep{sun+23} of continuously-driven, subsonic, shear flows and reported  sustained particle acceleration via second-order Fermi processes.

In the supersonic regime, the shear generally develops into multiple, irregular, small-scale shocklets. 
When a particle remains close to a shocklet, first-order Fermi processes lead to faster and arguably more efficient energization than in subsonic cases.
Acceleration in supersonic turbulence has received much less attention. The only instance of using kinetic simulations to study the problem that we are aware of is the very recent work by \citet{gootkin+25}, who performed hybrid simulations of decaying supersonic turbulence  and showed ion acceleration with a power-law energy spectrum  $\propto E^{-2.5}$.

In this paper, we focus on the decaying, transonic/supersonic shear that may arise in the presence of non-coherent bulk flows or as substructures in supersonic turbulence.
We use hybrid particles-in-cells (PIC) simulations with kinetic ions and fluid electrons to study how the free energy is channeled into accelerated ions, as a function of the Mach number of the shear flow.
In a companion paper (\citet{liang+25b}, henceforth Paper II), we discuss how pre-existing energetic particles may contribute to the evolution of the shear flow itself, inducing a CR viscosity \citep[][]{earl+88} that affects the momentum transfer and the partitioning of the free energy into heat, non-thermal particles, and magnetic turbulence.

This paper is organized as follows. The simulation setup and parameters are outlined in \S\ref{sec:setup}. 
In \S\ref{sec:benchmark} we validate the performance of the hybrid code by simulating a subsonic shear flow, from linear to nonlinear turbulent regimes. 
We dial the Mach number up in \S\ref{sec:tanh} and explicitly check for particle acceleration and magnetic field amplification. 
We discuss potential applications and open questions in \S\ref{sec:discussion} and conclude in \S\ref{sec:conclusion}.

\section{Simulation Setup} \label{sec:setup}
\begin{figure}
\centering
\includegraphics[width=\columnwidth]{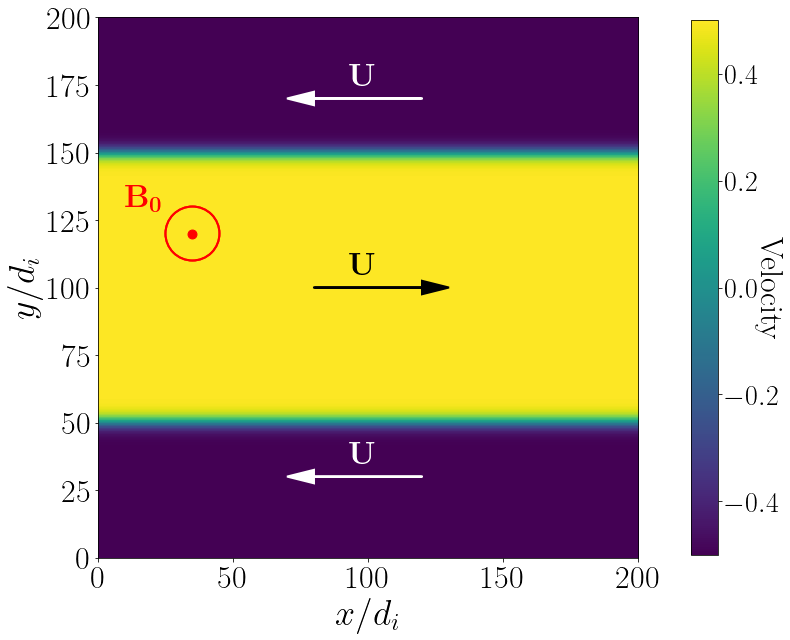}
\caption{Initial setup for the trans-Alfv\'enic and subsonic shear Run $\mathcal{B}$. 
The velocity shear is set by Equation \ref{eq:shear} and the initial magnetic field is mostly out-of-plane along $z$, with a small component along $x$ such that $B_{0,z}=20B_{0,x}$, following \citet{henri+13}. 
The layers at $50d_{\text{i}}$ and $150d_{\text{i}}$ have a width of 3$d_i$ and $\mathbf{B}\cdot (\nabla \times\mathbf{U})<0$ and $>0$, respectively.
}
\label{fig:Ux}
\end{figure}

We use the massively-parallel code {\tt\string dHybridR} \citep{gargate+07, haggerty+19a}, in which ions are kinetic macroparticles that evolve under the relativistic Lorentz force and electrons a massless, neutralizing, adiabatic fluid.
We start with a benchmark simulation of a subsonic/sub-Alfv\'enic shear and gradually increase the velocity to make the flow supersonic and super-Alfv\'enic. 

The system is two-dimensional (2D in the $x$–$y$ plane) and 3D in momentum space; 
the three components of the electromagnetic fields are also retained. 
Physical quantities are normalized to their initial values. 
The mass density is $\rho_0 \equiv m_i n_0$, where $m_i$ is the ion (proton) mass and $n_0$ is the ion number density.
The magnetic field is normalized to $B_0$, velocities to the initial Alfv\'en speed $\va \equiv B_0/\sqrt{4\pi \rho_{i,0}}$, and lengths to the ion inertial length $d_i \equiv c/\omega_{p}$, where $c=100 v_A$ is the speed of light and $\omega_{p}\equiv \sqrt{4\pi n_0 e^2/m_i}$ is the ion plasma frequency;
time is in unit of the inverse ion cyclotron frequency $\omega_{c}^{-1}\equiv m_ic/(eB_0)$. 
The thermal ion gyroradius equals the ion inertial length $d_i$, i.e., the thermal speed $v_{\rm th,i}=v_A$ and $\beta_i\sim 2$ unless otherwise specified.
Electrons are in thermal equilibrium with ions and have an adiabatic equation of state $P_e \propto \rho^{5/3}$ \citep{caprioli+18}. 

The Cartesian box has periodic boundary conditions in both $x$ and $y$, along and transverse to the flow, respectively.
Our default spatial domain measures $[L_x, L_y]$ with $L_x=L_y=L\equiv 200$, discretized on $N_x=N_y=1200$ grid points; 
$100$ particles per cell are used to ensure sufficient statistics and phase-space resolution. 
$\mathbf{B}_0$ is homogeneous and predominantly out-of-plane (along $z$, with a small component $B_x=B_z/20$) for analogy with \cite{henri+13}, who conducted a study of the evolution of the KHI with different fluid and kinetic approaches (\S\ref{sec:benchmark}).
The initial velocity field $\textbf{U}= U_{x}(y)\textbf{e}_{x}$ contains a double shear layer, with velocity varying from $-U_0$ to $+U_0$. 
The layers, with width $\Delta y = 3 d_i$, are located at $\tilde y_{1}=L_{y}/4$ and $\tilde y_{2}=3L_{y}/4$. 
The double tangential velocity shear profile $U_{x}(y)$ is (see Figure \ref{fig:Ux}):
\begin{equation}\label{eq:shear}
   \frac{U_{x}(y)}{U_0}\equiv \left[ \tanh\left(\frac{y-\tilde y_{1}}{\Delta y}\right)
    -\tanh\left(\frac{y-\tilde y_{2}}{\Delta y}\right)-1 \right]. 
\end{equation}
If $\mathbf{\Omega}\equiv\nabla \times \mathbf{U}$ is the vorticity, we see that the bottom/top shearing layers have $\mathbf{B}\cdot \mathbf{\Omega}< 0$ and $\mathbf{B}\cdot \mathbf{\Omega}>0$, respectively. 

Initial perturbations are not needed in particles-in-cells codes, since fluctuations arise naturally from the finite phase-space resolution. 
Space/time resolution and number of particles per cell have been chosen to ensure numerical convergence.

\begin{table}
\centering
\begin{tabular}{|c|c|c|c|c|c|c|}
\hline
Run & $M_{\rm A}$ & $v_{\text{th},i} [v_A]$ & $M_{\rm s}$ & $\beta_i$ & $L [d_i]$ & $\delta t [\omega_{c}^{-1}]$ \\
\hline
$\mathcal{B}$   & 1   & 1   & 0.55 & 2   & 200 & $2.5\times10^{-2}$ \\
\hline
$\mathcal{D}1$  & 2   & 1   & 1.1  & 2   & 200 & $2.5\times10^{-2}$ \\
$\mathcal{D}2$  & 4   & 1   & 2.2  & 2   & 200 & $2.5\times10^{-3}$ \\
$\mathcal{D}3$  & 8   & 1   & 4.4  & 2   & 200 & $1.25\times10^{-3}$ \\
$\mathcal{D}4$  & 16  & 1   & 8.8  & 2   & 200 & $6.25\times10^{-4}$ \\
$\mathcal{D}5$  & 20  & 1   & 10.9 & 2   & 200 & $5\times10^{-4}$ \\
\hline
$\mathcal{L}1$  & 8   & 1   & 4.4  & 2   & 100 & $1.25\times10^{-3}$ \\
$\mathcal{L}2$  & 8   & 1   & 4.4  & 2   & 400 & $1.25\times10^{-3}$ \\
\hline
$\mathcal{T}1$  & 4   & 2   & 1.1  & 8   & 200 & $2.5\times10^{-3}$ \\
$\mathcal{T}2$  & 4   & 4   & 0.5  & 32  & 200 & $2.5\times10^{-3}$ \\
$\mathcal{T}3$  & 4   & 8   & 0.3  & 128 & 200 & $2.5\times10^{-3}$ \\
$\mathcal{T}4$  & 8   & 2   & 2.2  & 8   & 200 & $2.5\times10^{-3}$ \\
\hline
\end{tabular}
\caption{Run parameters: shear maximum Alfv\'enic Mach number $M_{\rm{A}}$, ion thermal velocity $v_{\text{th},i}$, sonic Mach number $M_{\rm{s}}$, ion plasma $\beta_i$, box size $L$, timestep $\delta t$. 
Simulations are grouped according to the parameters varied: 
flow speed in $\mathcal{D}$, box size in $\mathcal{L}$, and temperature/plasma $\beta_i$ in  $\mathcal{T}$ runs.}
\label{tab:newsimparams}
\end{table}

Simulation runs are listed in Table \ref{tab:newsimparams} based on the parameter being varied: $\mathcal{D}$ for drift velocities (shear strength), $\mathcal{L}$ for domain sizes, $\mathcal{T}$ for particle thermal velocities and thus different plasma $\beta$.
We introduce the sonic and Alfvènic Mach numbers $M_{\rm{A}}\equiv 2U_0/v_A$ and $M_s\equiv 2U_0/c_s$, where the sound speed $c_s=\sqrt{2\gamma}v_{\rm th,i}$ is based on the total (ion + electron) pressure.
Hence, the benchmark Run $\mathcal{B}$ has $M_A=1$, $M_{\rm{s}}=0.55$, and $\beta_i=2$, i.e., its shear flow is trans-Alfv\'enic and subsonic. 

\section{Subsonic Shear}
\label{sec:benchmark}
Run $\mathcal{B}$ is chosen to match the benchmark of \citet{henri+13}, who compared  MHD, two-fluid, hybrid and PIC approaches.
Following their analysis, we single out three main stages of the dissipation of shear: the kinetic relaxation, the linear KHI growth, and a nonlinear turbulent regime when vortices merge to form secondary KHI and distort the flow into turbulence.

\subsection{Initial Kinetic Relaxation}
\begin{figure}
\centering
\includegraphics[width=\columnwidth]{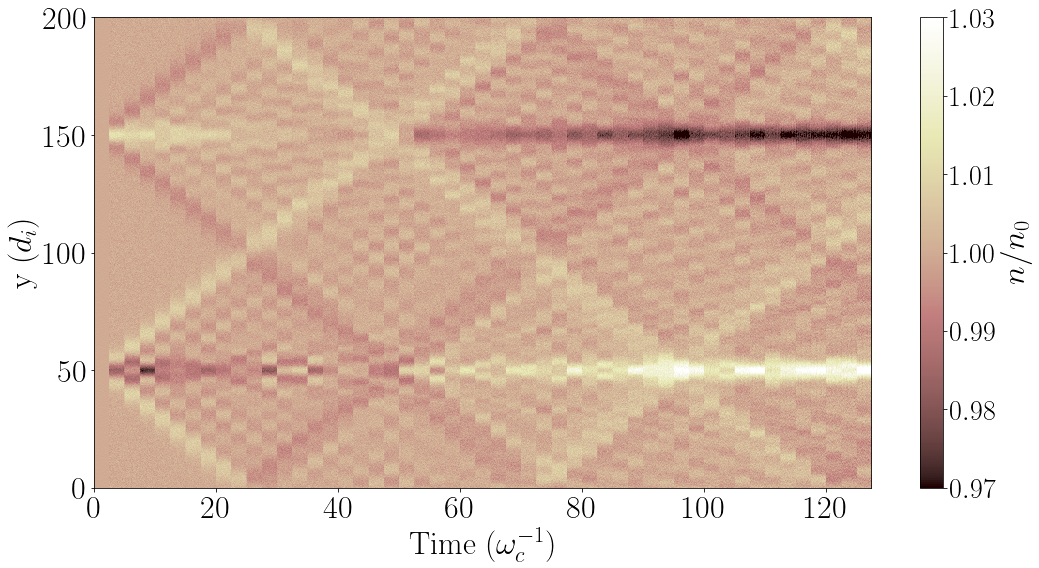}
\caption{Kinetic relaxation of the number density $n/n_{0}$ for the sub-Alfv\'enic and subsonic Run $\mathcal{B}$ in the first $125 \omega_{c}^{-1}$, before KHI kicks in. Disturbances in density propagate through the simulation domain at a constant speed.}
\label{fig:relax}
\end{figure}
Initially, the system is in a fluid but not Vlasov equilibrium \citep{cai+90}.
The slight tilt in the guide field modifies the $\mathbf{E}\times \mathbf{B}$ drift, producing a Lorentz force across the shear layers that drives a density inhomogeneity. 
Under such a force, at $t\lesssim 50\omega_{c}^{-1}$  the plasma develops a slight underdensity at $ \mathbf{B}\cdot\mathbf{\Omega}<0$ and an overdensity at $ \mathbf{B}\cdot\mathbf{\Omega}>0$, with amplitude $\delta n/n_{0}\sim 0.02$. 
A kinetic relaxation modifies the initial profile: 
as shown in Figure \ref{fig:relax}, sound waves propagate along $y$ and crosse the box several times due to the  periodic boundary conditions, eventually reversing the sign of the density fluctuations.
This phenomenon, which depends on the plasma temperature, does not affect the main results of this work.

\subsection{Linear Growth of Kelvin-Helmholtz Instability}
\begin{figure*}
\centering
\includegraphics[width=\textwidth]{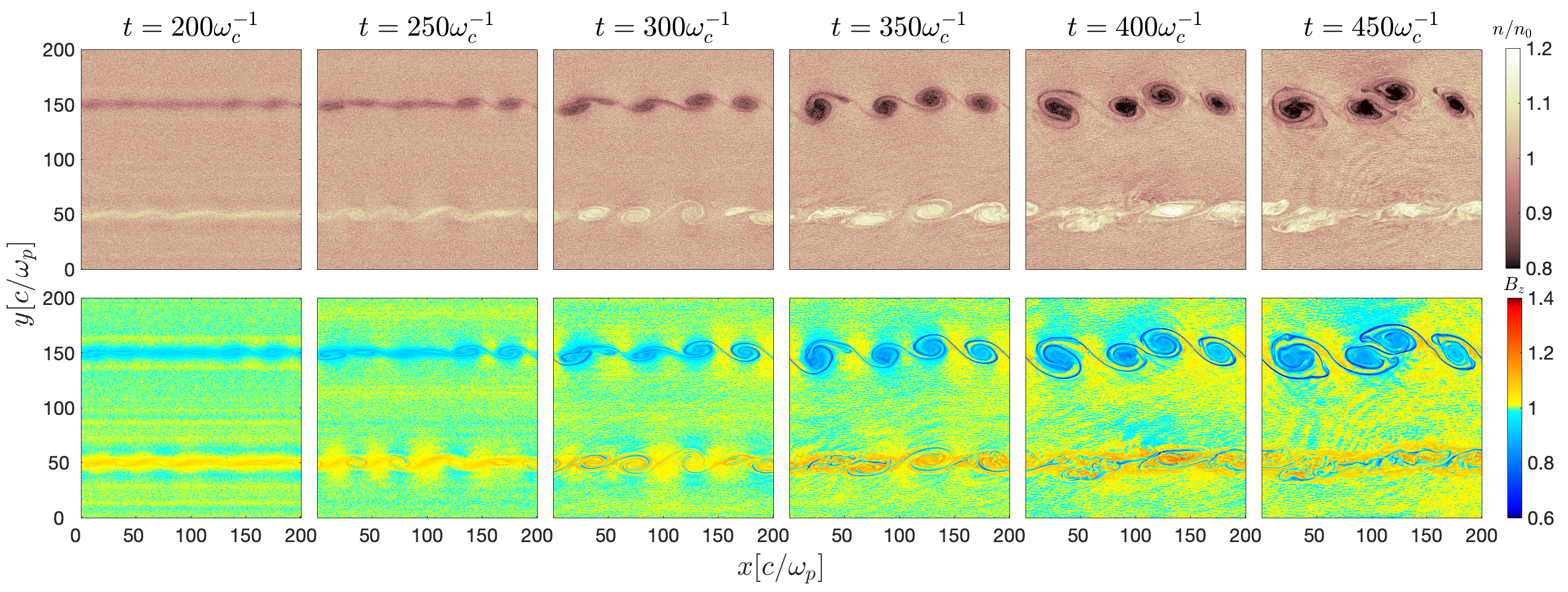}
\caption{Time evolution of out-of-plane magnetic field $B_{z}$ and density $n/n_{0}$ fore the subsonic simulation Run $\mathcal{B}$, showing the linear to the nonlinear evolution of the KHI, which for the chosen parameters has a growth time of $\sim 50 \omega_c$.}
\label{fig:suba}
\end{figure*}

Linear KHI  perturbations grow in amplitude  at both shearing layers until small vortices form at the contact surfaces. 
We estimate the growth rate of the fastest growing mode with $m=2$ (which corresponds
to a wavenumber $k_{\rm KH} d_i =1/30$) as $\gamma_{\text{KH}}\sim 0.015 \omega_{c}^{-1}$ at $ \mathbf{B}\cdot\mathbf{\Omega}<0$ and $\gamma_{\text{KH}}\sim 0.02 \omega_{c}^{-1}$ at $ \mathbf{B}\cdot\mathbf{\Omega}>0$, in agreement with \citet{henri+13}.
We also consistently find the growth of KHI to appear asymmetric at the two layers, and larger at the $ \mathbf{B}\cdot\mathbf{\Omega}>0$ layer; 
this is a kinetic effect due to a dependence of the growth rate on the sign of $\mathbf{B}\cdot\mathbf{\Omega}$, a feature previously observed in hybrid and full PIC simulations.

\subsection{The Turbulent Regime}
\label{subsec:non-lin}
In the nonlinear regime, secondary KH instabilities develop and vortices merge to generate structures at larger scales.
Figure \ref{fig:suba} shows that layers are still asymmetric, with magnetic fluctuations tracing the density fluctuations closely: under-dense regions are not sustained by magnetic pressure, but likely by turbulent pressure.
At $t=450 \omega_{c}^{-1}$, we also observe sound waves triggered by density fluctuations close to the lower overdense layer.

To capture how the initial free kinetic energy is dissipated, we introduce a parameter $\Delta(t)$ that quantifies the instantaneous amount of shear in the system. 
The momentum flux in the $x-$direction at time $t$ reads
\begin{equation}
    T_{xx}(t) = \iint \rho(x, y, t) v_x(x, y, t)^2\, \mathrm{d}x \, \mathrm{d}y,
    \label{eqn:Txx}
\end{equation}
where $\rho(x,y)$ and $v_{x}(x,y)$ are the ion mass density and average velocity along $x$ at position $(x,y)$; 
for supersonic flows along $x$, $T_{xx}$ is also proportional to the kinetic energy density.
Finally, the normalized momentum flux 
\begin{equation}
    \Delta(t)\equiv \frac{T_{xx}(t)}{T_{xx}(0)},
    \label{eqn:Delta}
\end{equation}
quantifies the fraction of shear kinetic energy/momentum remaining in the system at any time. 

\begin{figure}
    \centering
    \includegraphics[width=0.99\linewidth]{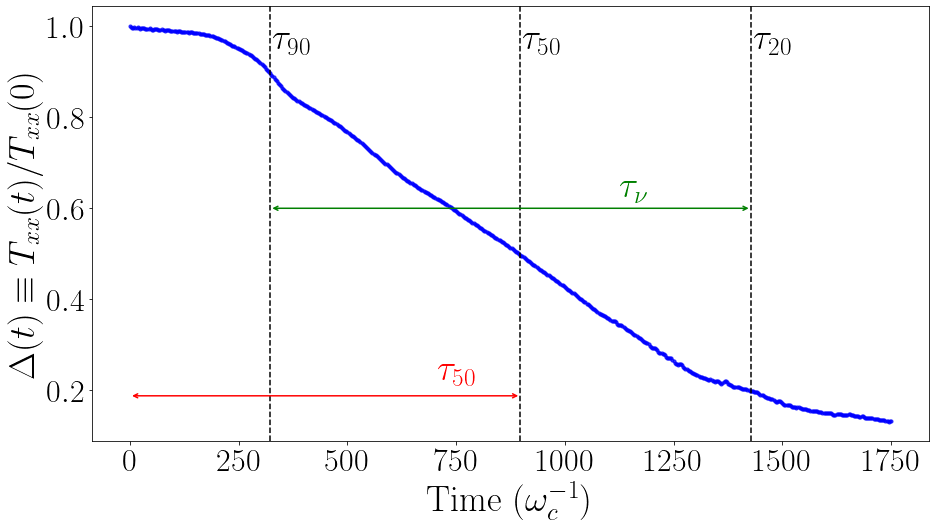}
    \caption{Evolution of $\Delta$ (Equation \ref{eqn:Delta}) for Run $\mathcal{B}$.
    $\tau_X$ is the time for $\Delta$ to reduce to $X\%$, so that $\tau_{90}$ measured the KHI onset timescales, $\tau_{50}$ gives a general estimate for the shear dissipation, and the viscosity timescale $\tau_{\nu}\equiv \tau_{20}-\tau_{90}$ determines to the duration of the nonlinear stage.}
    \label{fig:shear_t}
\end{figure}

The time evolution of $\Delta$ is presented in Figure \ref{fig:shear_t} for the benchmark  Run $\mathcal{B}$. 
If we introduce $\tau_X$ as the time for  $\Delta$ to reduce to $X\%$, we have three characteristic timescales, which will be used throughout the paper: 
(1) the \emph{halving time} $\tau_{50}$, which gives a general estimate for the shear dissipation;
(2)) the \emph{onset time} $\tau_{90}$, which measures the time that it takes for the collisionless shearing layers to couple due to either KHI or other kinetic instabilities (e.g., streaming instability if the shear is locally super-Alfv\'enic );
(3) the shear \emph{viscosity timescale} $\tau_{\nu}\equiv \tau_{20}-\tau_{90}$, which quantifies the duration of the nonlinear stage.

\section{Supersonic Shears}\label{sec:tanh}
We now discuss Runs $\mathcal{D}1$--$\mathcal{D}5$, where  the particle drift velocity is increased to supersonic values, while keeping all other plasma parameters the same. 

\subsection{Shear Dissipation}
\begin{figure*}
\centering
\includegraphics[width=\textwidth]{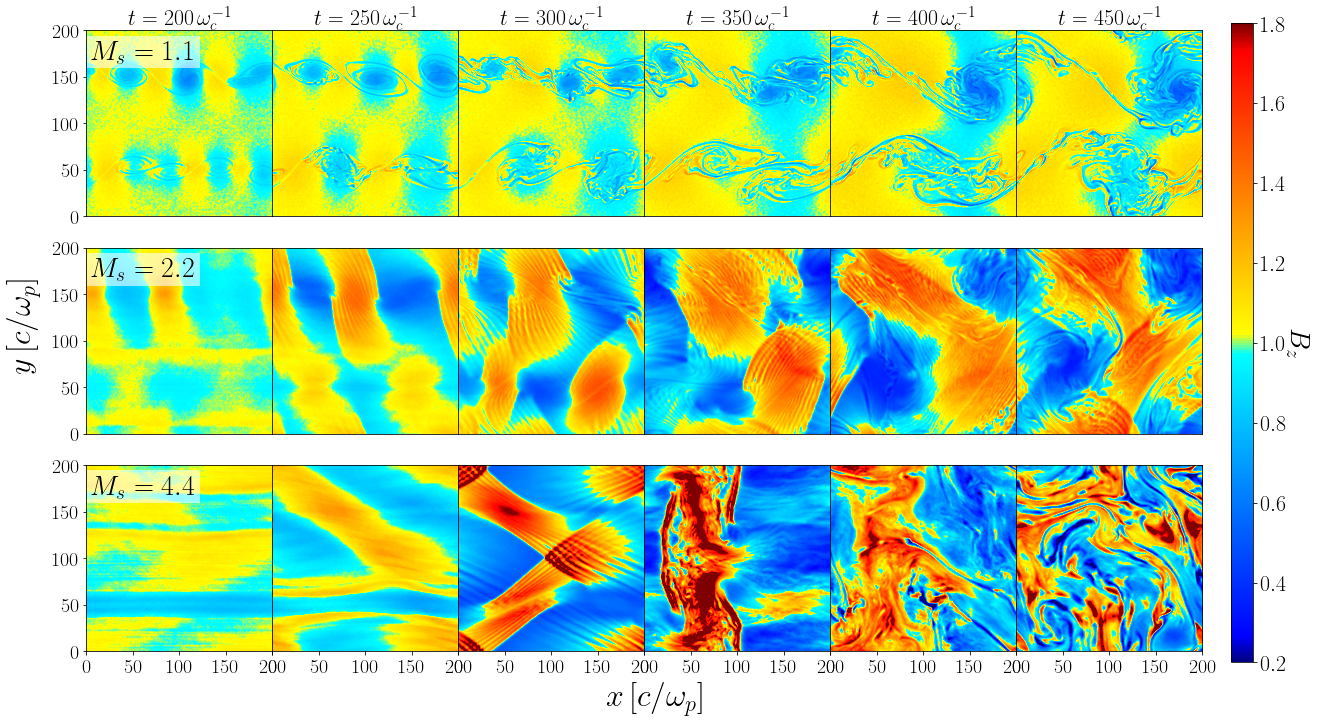}
\caption{Out-of-plane magnetic field $B_{z}$ over time for Run $\mathcal{D}1-3$ with $M_s=1.1, 2.2, 4.4$ as labeled, from transonic to supersonic. At higher $M_{\rm s}$ local shocklets grows to distort the initial velocity structure and evolved in turbulence.}
\label{fig:BzMs}
\end{figure*}

\begin{figure*}
\centering
\includegraphics[width=\textwidth]{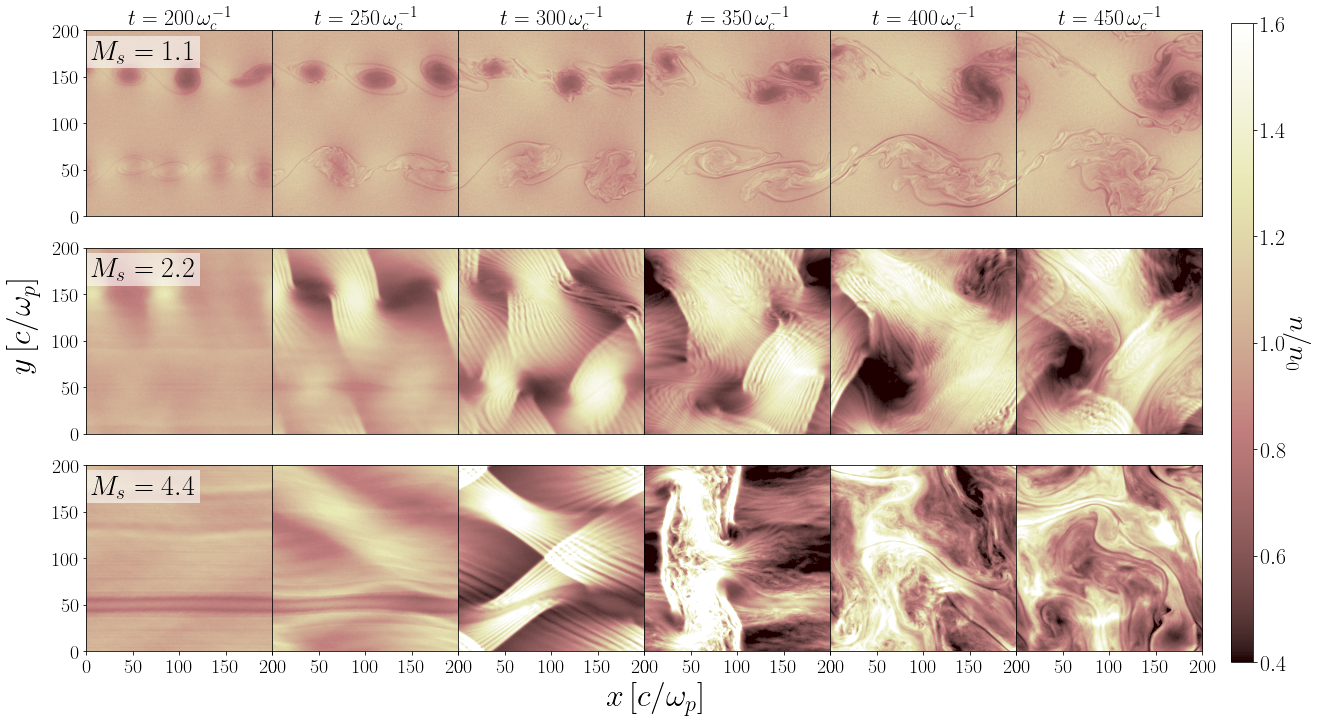}
\caption{As in Figure \ref{fig:BzMs}, but for density. Fluctuations follow the magnetic field patterns and become very prominent for supersonic shears.}
\label{fig:rhoMs}
\end{figure*}

Supersonic flows with $M_{\rm{s}}>1$ are compressible and the shear is dissipated in a different way.
Figures \ref{fig:BzMs} and \ref{fig:rhoMs} show the evolution of magnetic and density fluctuations as we dial up $M_{\rm{s}}$ for $\mathcal{D}1$ -- $\mathcal{D}3$. 
The transonic Run $\mathcal{D}1$ (top rows) appears quite similar to the subsonic case: 
KHI vortices are clearly observed, though they evolve, merge, and saturate faster than in Run $\mathcal{B}$, with fluctuations at larger amplitude.
The asymmetry in the growth rate of different shearing layers is also present. 
Already at $M_s=2.2$ (Run $\mathcal{D}2$, middle rows) the distinctive features of the KHI give way to more coherent and pronounced density and magnetic fluctuations, which we may label as \emph{shocklets}. 
Shocklets are more visible in Run $\mathcal{D}3$ with $M_{s}=4.4$ (bottom rows). 
At $t=300 \omega_{c}^{-1}$, steep gradients in field and density interact and form larger-scale structures, which in $\sim 100 \omega_{c}^{-1}$ become fully turbulent.
At even higher Mach numbers ($M_s=8.8, 10.9$, Runs $\mathcal{D}4-5$), supersonic ``waves" disrupt the shear almost immediately, which suggests that in realistic environments it is impossible to support such large velocity gradients over regions only a few $d_i$ thick.

\begin{figure}
    \centering
    \includegraphics[width=0.99\linewidth]{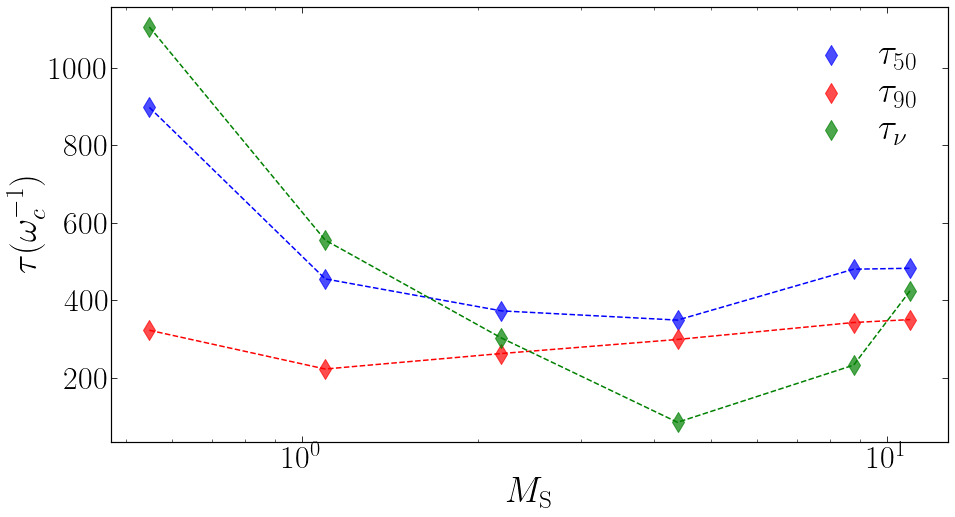}
    \caption{Characteristic dissipation timescales for Run $\mathcal{B}$ and $\mathcal{D}1 - \mathcal{D}5$ with different $M_{\rm s}$. 
    The shear-reducing time shortens as the flow becomes supersonic and saturates for $M_s\gtrsim 2.2$; since larger $M_{\rm s}$ runs have more free kinetic energy, this effectively corresponds to a larger dissipation rate.}
    \label{fig:tau_Ms}
\end{figure}

\subsection{Dissipation Timescales}
The natural timescale for the shear dissipation would be the eddy turnover (or crossing) time, $\tau_{\rm eddy} \sim L_{\rm{eddy}}/U_0$, with $L_{\rm{eddy}}\approx L$, but this may not be case at high $M_{\rm s}$, where shocklets appear, so we consider the three timescales defined above, instead.

Figure \ref{fig:tau_Ms} shows how the onset timescale $\tau_{90}$ is weakly dependent on $M_{\rm s}$.
Both $\tau_{50}$ and $\tau_{\nu}$ decrease with increasing $M_{\rm s}$ up to $M_{\rm{s}}\lesssim 4.4$, attesting to an overall faster dissipation of the shear across the subsonic to supersonic transition. 
At $M_{\rm s}\gtrsim 4$ the dissipation timescales remain comparable in cyclotron times, imply an effective dissipation \textit{rate} that increases for faster and faster shears, since they contain initial kinetic energy 
$\propto U_0^2$.

In supersonic and super-Alfv\'enic cases two distinct processes may take over the KHI and contribute to the coupling of the shears.
On one hand, beam/two-stream instabilities \citep[e.g.,][]{gary+84} can be triggered  when the relative velocity between two adjacent shearing layers exceeds $v_{A}$, producing both longitudinal and transverse fluctuations that destabilize the shear.
On the other hand, as we show below, the generation of nonthermal particles that scatter across the layer can effectively mediate energy and momentum transfer. 
This effective CR viscosity \citep{earl+88} accelerates the dissipation of shear, and will be further tested in Paper II, which also explores the conversion of the shear kinetic energy budget into heating, magnetic amplification, and nonthermal particles in detail.
The non-monotonic dependence of shear reducing timescales on $M_{\rm s}$ may suggest a non-trivial interplay between the two effects.

\begin{figure}
    \centering
    \includegraphics[width=0.99\linewidth]{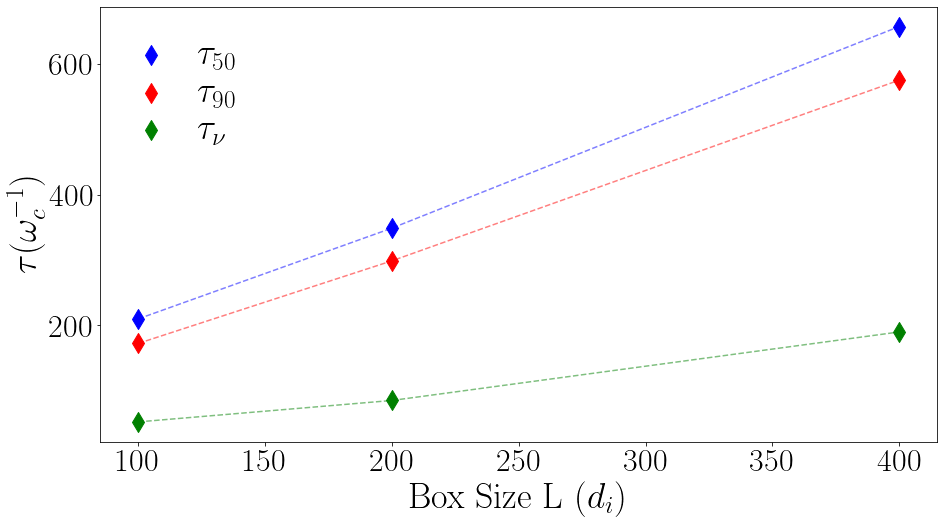}
    \caption{Characteristic timescales $\tau_{50}$, $\tau_{90}$, and $\tau_{\nu}$ for different box sizes $L$ for runs with fixed $M_s = 4.4$.  }
    \label{fig:box}
\end{figure}

\subsection{Dependence on the Box Size}
Runs ($\mathcal{D}1-5$) are performed in a fixed box of side $L=200d_i$ with different values of $U_0$. 
Yet, the shear depends not only on the velocity, but also on the spatial scale over which this change occurs. 
If we define the \textit{shear parameter} $\sigma \equiv M_s / L$, one might expect that runs with the same $\sigma$ would exhibit similar  behaviors. 

Figure \ref{fig:box} shows the shear dissipation timescales for runs with fixed $M_s = 4.4$ but different box sizes (Run $\mathcal{D}3$, $\mathcal{L}1$, and $\mathcal{L}2$).
We find that the characteristic timescales $\tau_{50}$ and $\tau_{90}$  increase with $L$, though less than linearly.
Moreover, run  $\mathcal{L}1$ and $\mathcal{L}2$ have the same $\sigma$ as $\mathcal{D}4$ and $\mathcal{D}2$, respectively, but for them  $\tau_{50}$ and $\tau_{90}$ scale proportionally to the box size $L$.
Finally, $\tau_{\nu}\propto \sqrt{\sigma}$ for fixed $M_{\rm s}$.
A possible explanation for these trends is that abating the shear takes longer in larger boxes because the ``viscous'' region (the shear layer) has a fixed width of $3d_i$ in our runs.
In general, the relaxation processes that govern $\tau_{\nu}$ depend on the competition between the available kinetic energy (larger at higher $M_{\rm s}$) and the dissipation rate (also faster at higher $M_{\rm s}$). 
We conclude that that the shear parameter $\sigma$ alone does not generally control the dissipation timescales, which rather depend on $M_{\rm s}$, $L$, and the layer thickness.
In Paper II we will consider sinusoidal supersonic shear profiles, which do not exhibit a sharp velocity change and may be more representative of astrophysical realizations.

\subsection{Acceleration Efficiency}
\begin{figure}
\centering
\includegraphics[width=0.99\columnwidth]{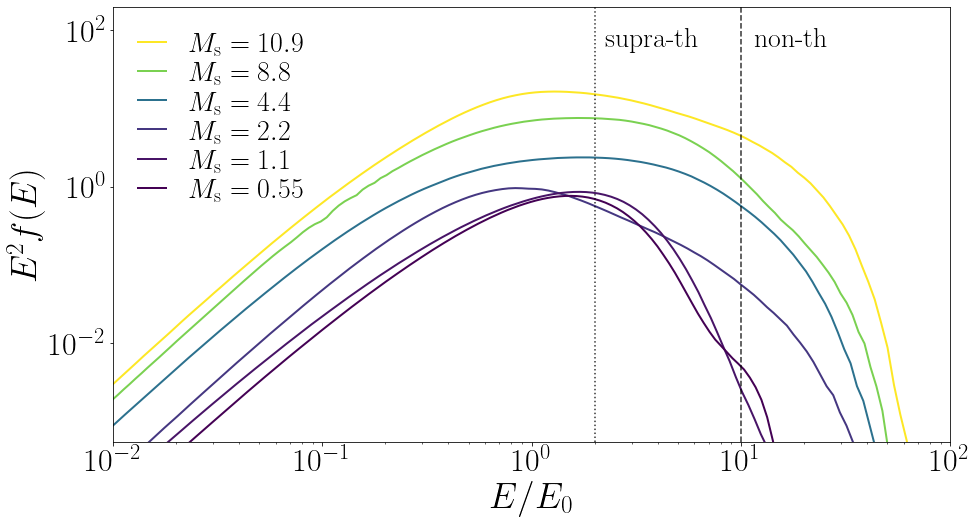}
\includegraphics[width=0.99\columnwidth]{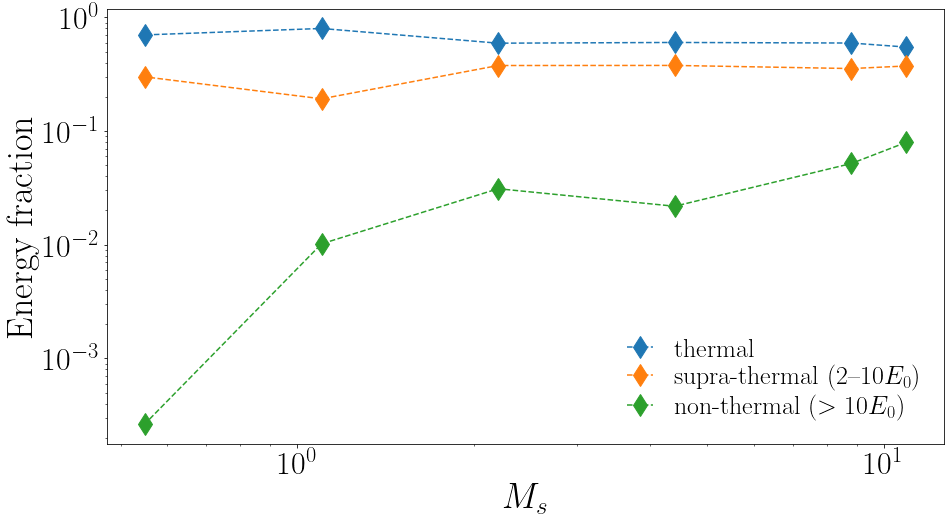}
\caption{Top panel: ion energy spectra for Run $\mathcal{B}$, $\mathcal{D}1 - \mathcal{D}5$ at their respective $\tau_{20}$, with $E_{0}$ defined in Equation \ref{eq:E0}. 
The thermal, suprathermal, and nonthermal regimes \citep{caprioli+15} are separated by vertical lines and contain the energy fractions shown in the bottom panel.
Note how the nonthermal acceleration efficiency increases with $M_{\rm s}$.
}
\label{fig:espec}
\end{figure}

Let us now consider the particle spectra from simulations with different $M_{\rm s}$.
The top panel of Figure \ref{fig:espec} shows the energy spectra extracted at their respective $\tau_{20}$, with energies normalized to their respective typical value $E_0$ defined by 
\begin{equation}\label{eq:E0}
    E_{0}\equiv \frac{1}{2}m_{i}U_0^2+\frac{3}{2}m_{i}v_{\rm th,i}^2,
\end{equation}
i.e., a combination of the bulk kinetic energy due to large-scale shear and the thermal energy.
While for subsonic and transonic cases the spectrum is well described by a Maxwellian, possibly with a small suprathermal bump, for supersonic cases we find that a nonthermal tail gradually develops and becomes flatter, also extending to larger energies. 

To better characterize the energy spectrum, we separate particles into three regimes following the definition usually adopted for shocks \citep{caprioli+15,johlander+21}: 
thermal for  $E<2E_{0}$, suprathermal for $2 E_{0}<E<10 E_{0}$, and nonthermal for  $E>10E_{0}$. 
The bottom panel Figure \ref{fig:espec} reports the energy fraction in nonthermal and supra-thermal particles calculated at $\tau_{20}$ for each simulation. 
Consistent with the spectra in Figure \ref{fig:espec}, larger $M_{\rm{s}}$ lead to larger acceleration out of the thermal pool. 
The supra-thermal fraction is always above $20\%$, attesting that even subsonic/transonic shears need kinetic (post-MHD) corrections to be described;
the nonthermal fraction is negligible for shears with $M_s\lesssim 2$, but ramps up from a few percent at $M_s=2$ to $\lesssim 10\%$ for the fastest shears, demonstrating that supersonic shears are efficient particle accelerators. 
The supra/nonthermal tails exhibit slopes slightly steeper than $E^{-2}$ for all the supersonic runs, interestingly comparable with the tails reported by \citet{gootkin+25} in hybrid simulations of supersonic turbulence.

We expect the acceleration efficiencies reported here to represent lower limits on the asymptotic ones that would be measured if the shear, and hence the turbulence, were continuously driven, as discussed by \citet{liu+25}.

\subsection{Maximum Energy}
\begin{figure}
\centering
\includegraphics[width=0.99\columnwidth]{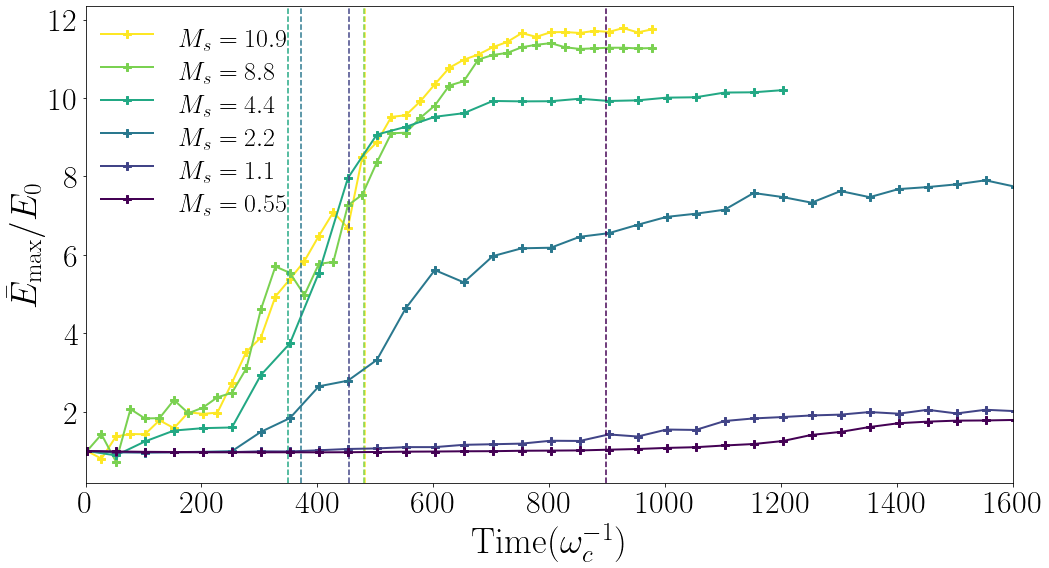}
\caption{Maximum particle energy over time for Run $\mathcal{B}$ and $\mathcal{D}1 - \mathcal{D}5$, normalized to their initial kinetic energy. Dashed lines indicate $\tau_{50}$ for different $M_{\rm{s}}$.
}
\label{fig:emax_weighted}
\end{figure}

We now examine the maximum energy obtained in each simulation.
Figure \ref{fig:emax_weighted} shows the time evolution of the weighted maximal particle energy
\begin{equation}
    \bar{E}_{\text{max}}\equiv\frac{\int E^{\text{n+1}}f(E) \,dE}{\int E^{\text{n}}f(E) \,dE}
\end{equation}
where $n\gtrsim 3$ is an integer number such that, for an the energy distribution $f(E)\propto E^{-m} \exp(-E/E_{\text{cut}})$, one obtains $\bar{E}_{\text{max}} \approx (n+1-m)E_{\text{cut}}$; 
following \cite{bai+15} we set $n=6$ and hence $\bar{E}_{\text{max}}\sim 5 E_{\text{cut}}$.
Different choices of $n$ just shift the overall trend, but do not change the hierarchy among different $M_{\rm s}$.

The maximum energy allowed in our setup is defined by the Hillas criterion  \citep{hillas84}, i.e., by the maximum potential drop associated with the motional electric field over the box size and reads $E_{\text{H}}\sim e U_0 B_0L/c$. 
We obtain $E_{\text{H}}\approx 40E_0$ for $M_{s}=10.9$ (Run $\mathcal{D}5$), consistent with the spectra in Figure \ref{fig:espec}, though the exponential cutoffs are lower by factors of a few.

Since there is minimal particle acceleration in the subsonic and transonic runs, $\bar{E}_{\text{max}}$ grows appreciably only when the flow becomes supersonic. The maximum energy increases rapidly and saturates on relatively short timescales, indicating a process with efficient energy gain. 
In Fermi processes the energy gain per cycle scales with the velocity of the scattering centers (possibly square for Fermi-II), but in our cases the acceleration rate is rather independent of the shear strength for $M_s\gtrsim 4.4$, which suggests that the interval between scatterings becomes longer for larger $M_{\rm s}$.

In any case, acceleration stops (sub-Hillas for most of the particles) because the shear is not driven and is disrupted on relatively short timescales $\tau_\nu\sim 200-500 \omega_c^{-1}$ (Figure \ref{fig:shear_t}). 
Again, driven simulations \citep{liu+25} suggest that ---if the shear is continuously regenerated--- acceleration to larger and larger energies should be achieved.

However, since spectra are steeper than $E^{-2}$, a truncation in the maximum attainable energy does not change much the energy budget in nonthermal particles.

\subsection{Magnetic Field Amplification}
\begin{figure}
\centering
\includegraphics[width=1\columnwidth]{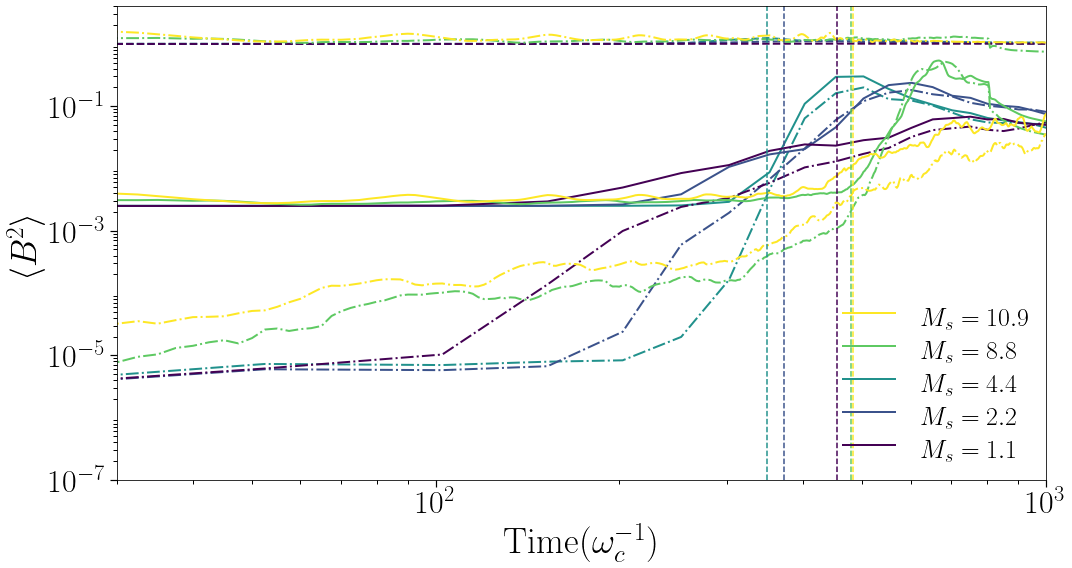}
\caption{Evolution of energy density in the components of the magnetic field for Runs $\mathcal{D}1-\mathcal{D}5$. Dashed lines indicate $\langle B_z^2 \rangle$, solid lines $\langle B_x^2 \rangle$, and dash-dot lines $\langle B_y^2 \rangle$. Vertical lines indicate $\tau_{50}$ of the simulations.
}
\label{fig:Bamp}
\end{figure}

In shearing flows magnetic fields can be amplified both by streaming instabilities \citep{kulsrud+68, bell04} and by turbulent dynamo \citep[e.g.,][]{moffatt78, tobias21}; 
also see \citet{achikanath-chirakkara+25} for very recent hybrid simulations of magnetic amplification via subsonic and supersonic turbulence in both weakly collisional and collisional plasmas.

Figure \ref{fig:Bamp} shows the growth of the three components of the magnetic fields, averaged over the box.
While the out-of-plane component $B_z$ remains quite unchanged, the in-plane components are appreciably amplified:  
$B_x$ starts from a finite value (section \ref{sec:setup}) and $B_y$ from the numerical noise, but they both peak at comparable values before damping, as expected for non-driven simulations. 

Interestingly, for $M_s\lesssim 4$ the amplification of such components starts later for larger $M_{\rm s}$, but they still peak at comparable timescales close to $\tau_{50}\sim 500 \omega_c^{-1}$. 
Amplification in $M_s\gtrsim 4$ cases, though, is more rapid and starts well before $\tau_{90}\sim 200\omega_c^{-1}$.
This suggests that for moderately supersonic shears most of the amplification comes from the turbulent dissipation of the shear layer, while for more supersonic shears there is also a contribution from the strong super-Alfv\'enic streaming drift. 
The change of slope in $B_y^2(t)$ observed at $t\sim \tau_{90}$ could thus be interpreted as the transition between the two regimes. 

Overall, the magnetic field amplification in our runs is of the order of $\delta B/B_0\lesssim 1$, with a true reorientation of the initial field only occurring for the most supersonic cases. 
It is worth stressing that, while magnetic field amplification slightly increases with $M_{\rm s}$, the peak magnetic energy is certainly not $\propto M_A^2$, negating the simple idea that a constant fraction of the free kinetic energy is converted into magnetic fields.
This also means that, unlike in small-scale turbulent dynamo theory, saturation is not controlled by the dynamical feedback of amplified magnetic fields that achieve equipartition with the kinetic components.
3D simulations, and with driven shears, may be necessary to capture the full picture of magnetic field amplification in supersonic shears. 

\subsection{High-Beta Plasmas}
\label{sec:highb}
\begin{figure}
\centering
\includegraphics[width=0.99\columnwidth]{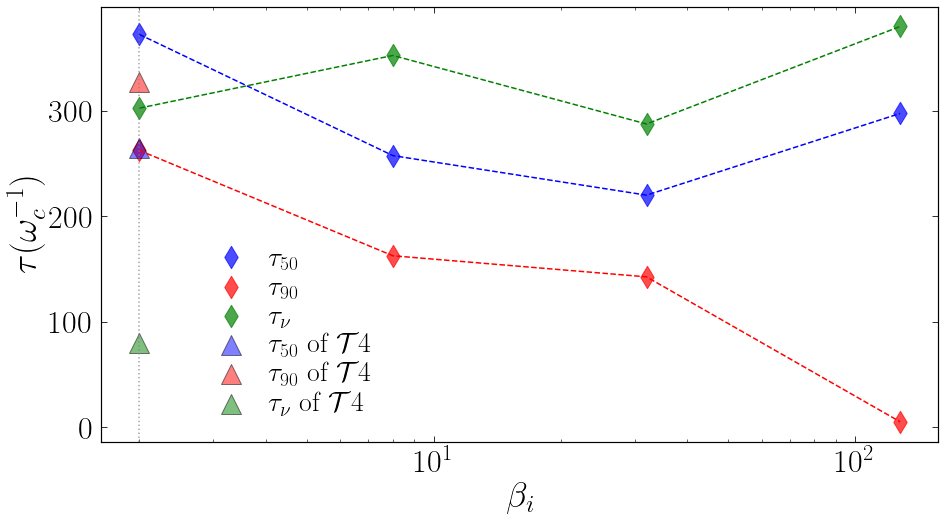}
\caption{Shear timescales for Runs with high $\beta_i$ (group $\mathcal{T}$ in Table \ref{tab:newsimparams}). 
Run $\mathcal{T}4$ with $M_s=2.2$ and $\beta_i=8$ is shown with triangles to contrast the other runs with a supersonic case in a high-$\beta$ plasma.}
\label{fig:highb}
\end{figure}

We now consider hot plasmas (high plasma $\beta_i$), with large sound speeds such that the shear flow may have $M_{\rm{A}}>1$ but $M_{\rm{s}}<1$. 
For these conditions, typical of the intracluster medium \citep[e.g.,][]{caprioli+19w} but also relatively common in the solar wind \citep[e.g.,][]{bale+09}, shears can be subsonic but super-Alfv\'enic.
In Runs $\mathcal{T}1-3$ we fix $M_{\mathrm{A}}=4$ but increase the plasma $\beta_i$, while in Run $\mathcal{T}4$ we keep $\beta_i=8$ and increase the particle drift velocity to have supersonic $M_s=2.2$.
Increasing the temperature to make the plasma incompressible suppresses the appearance of shocklets and the KHI develops its typical cat-eye vortices, similar to Run $\mathcal{D}1$. 
Figure \ref{fig:highb} shows the shear timescales in these runs: 
$\tau_{90}$ decreases monotonically, meaning that the dissipation starts earlier with larger thermal velocities, while $\tau_{50}$ and  $\tau_{\nu}$ are quite independent of $\beta_{i}$.
The supersonic Run $\mathcal{T}4$ exhibits a slightly longer $\tau_{90}$, but significantly smaller $\tau_{50}$ and $\tau_{\nu}$, consistent with the appearance of shocklets that drive a faster dissipation of the shear.

We note that in hot plasmas thermal particles have larger gyroradii $r_{\rm g, th}= v_{\rm th}/v_A d_i$, for our parameters possibly larger than the width, $3d_i$, of the velocity shear.
This allows particles to behave as long-range messengers that couple the shear flows and foster energy and momentum exchange, effectively increasing the viscosity and accelerating dissipation. 
A caveat is that realistic shear layers may only develop on scales comparable to $r_{\rm g, th}$, so that this effect is exaggerated in our simulations. 
We discuss extensively the role that energetic particles with large $r_{\rm g, th}$ have on shears in Paper II. 

In summary, in high-$\beta$ plasmas there are two competing effects that may explain why the dissipation timescales is essentially independent of the ion temperature:
on one hand, reducing the compressibility suppresses the shocklets that can be effective in disrupting the shear, while, on the other hand, the presence of particles with larger gyroradii may favor energy transport across the shear. 
Such finite-Larmor effects are not captured in fluid or MHD approaches but may be crucial to model viscosity in astrophysical plasmas.

\section{Astrophysical Applications}
\label{sec:discussion}
\subsection{Pressure Support in Molecular Clouds}

Shear layers in astrophysical environments, like the warm ISM, can experience runaway radiative cooling. As the gas radiates, its temperature drops. Under isobaric conditions, the density increases $n \propto T^{-1}$, which accelerates the cooling rate, $\Lambda \propto n^2$.  
The falling temperature also reduces the sound speed ($c_s \propto T^{1/2}$), often making initially subsonic flows supersonic ---a state common in the ISM and ICM.

Our simulations show that supersonic flows can generate shocklets that reheat the gas, potentially halting the thermal instability, and that tens/few percent of the free energy is generically converted into suprathermal/nonthermal ions, which are intrinsically less coupled to the thermal electrons, when a two-temperature plasma develops \citep[][]{kawazura+18}.
Also, transonic and mildly-supersonic shear can amplify the initial magnetic field, channeling $\lesssim 20\%$ of the kinetic energy into magnetic turbulence for $M_A\gtrsim 1$, effectively storing pressure in a form that cannot be radiated away.
Such non-thermal components, viz CRs and magnetic fields, do not radiate efficiently and can provide support against compression, even if the gas is thermally supersonic.
They may also account for the non-thermal support inferred in molecular clouds \citep[e.g.,][and references therein]{lin+25}. 

The ultimate contribution of such populations to cooling ISM/ICM patches would require a fully-kinetic description to capture both the collisional (Coulomb) and collisionless (mediated by collective electromagnetic forces) ion--electron coupling, also including a prescription for the radiative cooling. 
While extremely interesting and potentially pivotal for our understanding of the ISM and ICM, we are not aware of any first-principle calculations that can provide the final answer to this problem.

\subsection{CR Viscosity in Disks}
The arising of a nonthermal population may be important also for the viscous transport of angular momentum in accretion disks.
Such a transport is often thought to be mediated by magnetic turbulence, typically driven by the magneto-rotational instability \citep[MRI,][]{balbus+98};
the potential role of CRs has gathered much less attention \citep[e.g.,][]{kuwabara+14}, besides their contribution to the ionization of clouds and proto-planetary disk \citep[e.g.,][]{glassgold+12}.
The analytical calculations by \citet{earl+88} remain a milestone in the estimate of CR viscosity, but rely on a-priori assumptions for the CR spectrum and mean-free path.
While they conclude that for Galactic parameters the effects of CR viscosity are small, more work is needed to assess what happens when energetic particles are spontaneously produced by the shear itself, as shown above, or by other processes operating in disks.
We argue that the local enhancements of energetic particles and the scattering provided by self-generated magnetic turbulence may drastically increase the CR viscosity.
Kinetic simulations of disks have shown evidence of particle acceleration through various mechanisms \citep[e.g.,][]{riquelme+12, hoshino+15, kunz+16, bacchini+22, sandoval+24} but the relative contribution of MRI and viscosity induced by fully-evolved CR distributions in realistic environments has not been assessed, yet.

Though likely not important for Galactic CRs \citep[where nonetheless some diffusive reacceleration may be required,][]{drury+17}, disks may host CR seeds accreted from the ISM, possibly in energy equipartition with the thermal plasma. 
The contribution of such preexisting CRs is usually not taken into account in kinetic approaches, and it will be the subject of Paper II.

\section{Conclusions}
\label{sec:conclusion}
We performed hybrid simulations of decaying sub/supersonic shear flows, studying how quickly the shear is dissipated and how the free kinetic energy is channeled into energetic particles and magnetic turbulence, as a function of the shear sonic and Alfv\'enic Mach numbers (Table \ref{tab:newsimparams}). 
Our main results are the following: 

\begin{enumerate}
\item The time needed to dissipate the initial free energy decreases  from subsonic to supersonic flows and saturates for $M_s\gtrsim 4$ (Figure \ref{fig:tau_Ms}).
Still, since for larger shear velocities there is more free energy, this means that the dissipation rate increases with $M_{\rm s}$. 
The faster dissipation is driven by the appearance of shocklets, which replace the canonical KHI vortexes of the subsonic cases.

\item Supersonic shear flows are accompanied by significant particle acceleration. 
Ion distributions deviate more and more significantly from Maxwellians when $M_{\rm s}$ increases beyond 1 (Figure \ref{fig:espec}): they develop a suprathermal shoulder that contains $\sim 20\%$ of the total energy and a tail of nonthermal particles with efficiency up to $\lesssim 10\%$.
Similar conclusions were reached in the first kinetic study of decaying supersonic turbulence with hybrid simulations \citep{gootkin+25}.

\item The maximum energy achieved by accelerated ions also increases with $M_s\gtrsim 2.2$.
While few highest-energy particles in the simulations reach the Hillas limit, the bulk of the distribution cuts off at the energy achieved  around $\tau_{50}$, i.e., when half of the initial kinetic energy is dissipated (Figure \ref{fig:emax_weighted}).

\item Rearrangement and amplification of the initial magnetic field are always observed and become more prominent for large $M_{\rm s}$ (Figure \ref{fig:Bamp}).
They are due both to turbulent dynamo and super-Alfv\'enic streaming instability, with the latter contributing in the early stages of the most supersonic cases. 

\item Since in our simulations the shear is initialized and let decay, the maximum ion energy stalls after $\tau_{50}$ and the amplification of the magnetic field saturates at levels for which it remains dynamically unimportant; we do not expect this be the case if the shear were driven.

\item Increasing the plasma $\beta_i$ makes the dissipation faster (Figure \ref{fig:highb}), though the effect saturates if the flow becomes subsonic.
We ascribe this to the interplay between two competing effects: the reduced compressibility that suppresses the shocklets and the viscosity provided by particles with gyroradii exceeding the shear layer thickness.
\end{enumerate}

While far from presenting a comprehensive theory of the dissipation processes relevant in astrophysical supersonic shears and turbulence, our results generally suggest the need to go beyond fluid approaches to properly capture the potentially relevant effects of energetic particles and amplified magnetic fields.
In a companion paper we will analyze in more detail the role of both spontaneously-produced and preexisting energetic particles on the shear dynamics. 

We considered here only decaying shears, but recent MHD-PIC simulations show that ---if shear keeps being driven--- ion acceleration can proceed to larger energies via second-order Fermi processes \citep{liu+25}.   
The two approaches are highly synergistic, because hybrid simulations are needed to quantify the fraction of ions that can be injected into the acceleration process, while MHD-PIC ones can study the problem in significantly larger boxes, and potentially in global environments.
Taken together these simulations confirm that shear-driven turbulence can accelerate particles efficiently across a wide parameter space, from sustained, quasi-stationary turbulence to freely decaying, supersonic flows dominated by shocklets.

\begin{acknowledgments}
We would like to thank Thomas Berlok for his help in setting the initial conditions and calculating the linear growth rates of the KHI, and Ellen Zweibel, Peng Oh, Mateusz Ruszkovski, Anatoly Spitkovksy, Mingxuan Liu, and Xiaochen Sun for interesting discussion on CR acceleration in shearing layers.
We would like to also thank The University of Chicago Research Computing Center for providing the computational resources to conduct this research.
This research was partially supported by NASA grant 80NSSC18K1726, NSF grants AST-2510951 and AST-2308021 to D.C., and by NSF grant PHY-2309135 to the Kavli Institute for Theoretical Physics. N.L. acknowledges support from NSF grant AST-240752.
\end{acknowledgments}

\bibliography{main}
\bibliographystyle{aasjournalv7}
\end{document}